\documentclass[11pt]{article}
\usepackage{kotex}
\usepackage{amsmath}
\usepackage{amsthm}
\usepackage{graphicx}
\usepackage{multirow}
\usepackage{subcaption}
\usepackage{amssymb}
\usepackage{algorithm}
\usepackage{algpseudocode}
\usepackage[hidelinks]{hyperref}
\usepackage{breakurl}

\newtheorem{theorem}{Theorem}

\algtext*{EndFor}
\algtext*{EndFunction}

\title{Stable Marriage with One-Sided Preference}
\author{Seongbeom Park \\ \href{mailto:sparkamita90@gmail.com}{sparkamita90@gmail.com}}

\begin{document}
\maketitle

\begin{abstract}
    Many countries around the world, including Korea, use the school choice lottery system.
    However, this method has a problem in that many students are assigned to less-preferred schools based on the lottery results.
    In addition, the task of finding a good assignment with ties often has a time complexity of NP, making it a very difficult problem to improve the quality of the assignment.

    In this paper, we prove that the problem of finding a stable matching that maximizes the student-oriented preference utility in a two-sided market with one-sided preference can be solved in polynomial time, and we verify through experiments that the quality of assignment is improved.
    The main contributions of this paper are as follows.
    We found that stable student-oriented allocation in a two-sided market with one-sided preferences is the same as stable allocation in a two-sided market with symmetric preferences.
    In addition, we defined a method to quantify the quality of allocation from a preference utilitarian perspective.
    Based on the above two, it was proven that the problem of finding a stable match that maximizes the preference utility in a two-sided market with homogeneous preferences can be reduced to an allocation problem.
    In this paper, through an experiment, we quantitatively verified that optimal student assignment assigns more students to schools of higher preference, even in situations where many students are assigned to schools of low preference using the existing assignment method.
\end{abstract}

\section{Introduction}
In many countries around the world, including Korea and the United States, the first application, second choice method, which allows students to choose the school of their choice, is mainly used when assigning students to schools without testing.
Students create and submit a list of schools they wish to attend, and once the application is completed, the placement officer assigns students through a lottery according to the number of students at each school.
In this process, it is not possible to assign all students to the school they want, but students assigned to a school they do not want have a variety of problems, such as a lack of motivation to learn or difficulty adapting to the school due to the long commuting distance.

In general, the problem of assigning students is reduced to a \textit{stable marriage problem} and the \textit{Gale-Shapley algorithm}\cite{gale1962college} is used.
This is a method of creating matches based on preferences for each other in a \textit{two-sided market}, ensuring the \textit{stability} of the generated matches.
However, since the Gale-Shapley algorithm does not allow ties in the preference list, if there is a tie, the tie must be randomly removed through \textit{Tie-Breaking}\cite{irving1994stable}.
As a method to eliminate tied scores, a method is used to assign random numbers to students through a random drawing and assign students starting with the highest number.

Using the Gale-Shapley algorithm produces \textit{stable matching}, but the number of students assigned to each application varies greatly depending on the lottery results\cite{abdulkadiroglu2021school}.
This is because there may be multiple stable matches that can be generated from a given preference list\cite{irving1987efficient}.
The following example shows how much the quality of assignment varies.
Figure~\ref{fig:one_sided_pref} shows the preference list of three students.
Student A prefers schools A, B, and C in that order, while Students B and C prefer two schools at the same time as first and the other school as second.
Let's assume that every school has one student.
If the matching of Figure~\ref{fig:optimal_match} is assigned as \{(a, A), (a, B), (c, C)\}, all three students will be assigned to their first preferred school. .

\begin{figure}[h]
    \begin{subtable}{0.4\textwidth}
        \centering
        \begin{tabular}{c c c c}
            a: & A     & B & C \\
            b: & (A B) & C & ( ) \\
            c: & (B C) & A & ( )
        \end{tabular}
        \caption{Student's preference}
        \label{fig:one_sided_pref}
    \end{subtable}
    \hfill
    \begin{subtable}{0.6\textwidth}
        \centering
        \begin{tabular}{| c || c c c |}
            \hline
            \multirow{2}{*}{Matching} & \multicolumn{3}{c|}{Ranking} \\
                    & 1st & 2nd & 3rd \\
            \hline
            \hline
            \{(a, A), (b, B), (c, C)\} & 3 & 0 & 0 \\
            \hline
        \end{tabular}
        \caption{Student-optimal matching}
        \label{fig:optimal_match}
    \end{subtable}
    \caption{Student-optimal matching with student preference only}
\end{figure}

If strictly ordered preferences are required at the application stage, students b and c must apply separately to the two schools they equally prefer.
In a situation like Figure~\ref{fig:strict_student_pref}, where students apply based on preference with ties eliminated, and where the school gives equal preference to all students, as shown in Figure~\ref{fig:strict_school_pref}, there are two \textit{student-oriented} stable matching depends on the assignment order for students a and b.
Figure~\ref{fig:strict_alloc} shows the number of students assigned to each application based on matching.
Among the matches generated under the condition in which ties are eliminated, the ideal match of Figure~\ref{fig:optimal_match} is not included.

\begin{figure}[h]
    \centering
    \begin{subtable}{0.4\textwidth}
        \centering
        \begin{tabular}{c c c c}
            a: & A & B & C \\
            b: & A & B & C \\
            c: & B & C & A
        \end{tabular}
        \caption{Student's preferences}
        \label{fig:strict_student_pref}
    \end{subtable}
    \hfill
    \begin{subtable}{0.4\textwidth}
        \centering
        \begin{tabular}{c c}
            A: & (a b c) \\
            B: & (a b c) \\
            C: & (a b c)
        \end{tabular}
        \caption{School's preferences}
        \label{fig:strict_school_pref}
    \end{subtable}
    \hfill
    \begin{subtable}{\textwidth}
        \centering
        \begin{tabular}{| c || c c c |}
            \hline
            \multirow{2}{*}{Matching} & \multicolumn{3}{c|}{Ranking} \\
            & 1st & 2nd & 3rd \\
            \hline
            \hline
            \{(a, A), (b, C), (c, B)\} & 2 & 0 & 1 \\
            \{(a, C), (b, A), (c, B)\} & 2 & 0 & 1 \\
            \hline
        \end{tabular}
        \caption{Possible student-oriented matchings}
        \label{fig:strict_alloc}
    \end{subtable}
    \caption{Possible matchings with student's strictly ordered preference}
\end{figure}

In a situation where multiple schools are supported for each application, as shown in Figure~\ref{fig:tie_student_pref}, and schools equally prefer students, as shown in Figure~\ref{fig:tie_school_pref}, there are 4 types of stable matchings is created according to the students' assignment order.
Figure~\ref{fig:tie_alloc} shows the number of students assigned to each application according to matching.
Depending on which students are assigned first, all students may be assigned to the school of their choice, or some students may be assigned to a school with less preference.

\begin{figure}[h]
    \centering
    \begin{subtable}{0.3\textwidth}
        \centering
        \begin{tabular}{c c c c}
            a: & A     & B & C   \\
            b: & (A B) & C & ( ) \\
            c: & (B C) & A & ( )
        \end{tabular}
        \caption{Student's preferences}
        \label{fig:tie_student_pref}
    \end{subtable}
    \hfill
    \begin{subtable}{0.3\textwidth}
        \centering
        \begin{tabular}{c c}
            A: & (a b c) \\
            B: & (a b c) \\
            C: & (a b c)
        \end{tabular}
        \caption{School's preferences}
        \label{fig:tie_school_pref}
    \end{subtable}
    \hfill
    \begin{subtable}{\textwidth}
        \centering
        \begin{tabular}{| c || c c c |}
            \hline
            \multirow{2}{*}{Matching} & \multicolumn{3}{c|}{Ranking} \\
            & 1st & 2nd & 3rd \\
            \hline
            \hline
            \{(a, A), (b, B), (c, C)\} & 3 & 0 & 0 \\
            \{(a, A), (b, C), (c, B)\} & 2 & 1 & 0 \\
            \{(a, B), (b, A), (c, C)\} & 2 & 1 & 0 \\
            \{(a, C), (b, A), (c, B)\} & 2 & 0 & 1 \\
            \hline
        \end{tabular}
        \caption{Possible student-oriented matchings}
        \label{fig:tie_alloc}
    \end{subtable}
    \caption{Possible matchings with tie-breaking}
\end{figure}

The reason it is difficult to create an ideal match in the example situations is because constraints are added in the process of eliminating ties\cite{erdil2008s}.
Because Tie-Breaking does not consider the outcome of allocation, the constraints added according to the results of the random drawing greatly affect the quality of allocation.
In order to find the best quality allocation method, studies have been conducted from perspectives such as \textit{egalitarian} and \textit{minimum regret}.
If there is a tie, the time complexity is often Non-deterministic Polynomial time (NP)\cite{abdulkadiroglu2021school, manlove2002hard, o2007algorithmic},
Methods to improve the quality of assignments are being studied\cite{erdil2008s, irving2010finding, ravindranath2021deep}.

In this paper, we will prove that in a two-sided market with one-sided preference and the other side has limited capacity, the student-oriented stable matching that maximizes the utility \textit{student-optimal matching} can be found in polynomial time.
And through experiments, we will verify that the allocation quality is greatly improved.
The main contributions of this paper are as follows.
(1) It shows that the problem of finding a stable student-oriented match in a two-sided market with one-sided preference is the same as the problem of finding a stable match in a two-sided market with symmetric preferences.
(2) Quantify the quality of matching by defining utility from a preference utilitarian perspective.
(3) We propose a method to find a stable match that maximizes the preference utility in polynomial time, and prove that the match found by the proposed method is the student-optimal matching.
(4) Through experiments, we show how student-optimal matching improves assignment quality.

\section{The Assignment Criteria}
\subsection{Stable matching}
A stable match is defined as a match without \textit{blocking pairs}\cite{gale1962college}.
A blocking pair is one in which a pair that does not exist in the match prefers each other over the current partner, thus breaking up existing pairs.
In a two-sided market with a tie, stability extends to three types (weakly, strongly, super) depending on whether equally favored pairs are viewed as blocking pairs\cite{irving1994stable}.
In strong/super stable matching, a pair is considered to be a blocking pair even if it is preferred to the same degree, but in \textit{weakly stable matching}, it is viewed as a blocking pair only if it is preferred to a greater degree.
In this paper, we aim to find a weakly stable match that can assign all students.

\subsection{Student-oriented}
The Gale-Shapley algorithm generates matches that are advantageous to the applicant\cite{irving1987efficient, abdulkadirouglu2005new}.
In order to create a favorable allocation for students, students apply in order, starting with the schools they prefer most.
Since the school gives equal preference to all students, it assigns students to the school first in order of assignment and approves them until all seats are filled.
As a result, student-oriented matching assigns students who applied with a higher rank to a school with priority over students who applied with a lower rank.

A situation in which a student should be assigned priority over another student can be expressed as a school's preference.
Figure~\ref{fig:sym_inst} shows a two-sided market in which students' preferences for schools coincide with schools' preferences for students.
Because it was reorganized to create student-oriented matching through school preferences, the stable matching created from these preferences is always student-oriented.

\begin{figure}[h]
    \centering
    \begin{subtable}{0.3\textwidth}
        \centering
        \begin{tabular}{c c c c}
            a: & A     & B & C   \\
            b: & (A B) & C & ( ) \\
            c: & (B C) & A & ( )
        \end{tabular}
        \caption{Student's preferences}
        \label{fig:sym_student_pref}
    \end{subtable}
    \hfill
    \begin{subtable}{0.3\textwidth}
        \centering
        \begin{tabular}{c c c c}
            A: & (a b) & c & ( ) \\
            B: & (b c) & a & ( ) \\
            C: & c & b & a
        \end{tabular}
        \caption{School's preferences}
        \label{fig:sym_school_pref}
    \end{subtable}
    \caption{An instance with symmetric preferences}
    \label{fig:sym_inst}
\end{figure}

In Figure~\ref{fig:sym_inst}, if $r(i, j)$ is a function representing the ranking in which $i$ prefers $j$, then $r(s, h) = r(h, s)$ holds.
For example, the rank in which student a prefers school A ($r(\text{a}, \text{A}) = 1$) is the rank in which school A prefers student a ($r(\text{A}, \text{a}) = 1$).
When preferences match like this, it is called symmetric preference\cite{gai2007acyclic}.

\subsection{Maximizing utility}
John Harsanyi\cite{harsanyi1977morality} said ``\textit{preference utilitarianism} is the only form of utilitarianism consistent with the important philosophical principle of \textit{preference autonomy}.''
Under the principle of preference autonomy, deciding what is good or bad depends solely on the individual's preferences.
It would be best if we could satisfy all students, but if more students apply to a school than there are available seats, some students will inevitably fail.
When there are conflicting preferences among members of society, preference utilitarianism can be applied by assigning a certain weight to individual preferences and finding the result that provides the highest level of satisfaction\cite{singer2013companion}.
In this paper, given that the preference ranking of the assigned school had a significant impact on student satisfaction, the difference in weights according to the preference ranking was set to be large.

The preference utility $Q$ of matching $M$ is defined as $(|\mathbb{S}| + 1)$-adic number for the student set $\mathbb{S}$.
Each digit $q_k$ of the utility $Q(M)$ refers to the number of pairs in matching $M$ where the assigned partner's preference rank is $k$, and $z$ is the maximum value of the student's preference list length.

\begin{center}
    $Q(M) = \sum\limits_{k=1}^{z}{q_k (|\mathbb{S}| + 1)^{z - k}}$
\end{center}

\begin{center}
    $q_k = \lvert \{(s, h) \in M \mid r(s, h) = k\} \rvert ,\  0 \le q_k \le |\mathbb{S}|$
\end{center}

\begin{center}
    $z = \max\limits_{s \in \mathbb{S}, h \in \mathbb{H}}{r(s, h)}$
\end{center}

For example, in Figure~\ref{fig:tie_alloc}, $Q(\text{\{(a, A), (b, B), (c, C)\}}) = 300_{(4)}$ has the largest utility,
$Q(\text{\{(a, A), (b, C), (c, B)\}}) = Q(\text{\{(a, B), (b, A), (c, C)\}}) = 210_{(4)}$ have the same utility,
and $Q(\text{\{(a, C), (b, A), (c, B)\}}) = 201_{(4)}$ has the smallest utility.

The time complexity of the problem of finding a stable match that maximally satisfies egalitarianism or minimum regret in a two-sided market with symmetric preferences is known as NP-hard\cite{o2007algorithmic}.
In this paper, we find stable matching that maximizes the preference utility in a two-sided market with symmetric preferences.

\section{Stabale Marriage and Assigment Problem}
In this section, we will find student-optimal matching in a two-sided market where only the set of students $\mathbb{S}$ has preferences and the set of schools $\mathbb{H}$ has only a limited number of seats.
Student-optimal matching refers to the matching with the maximum utility among student-oriented stable matchings.
If a situation where only one side has a preference is transformed into a situation with symmetric preferences, stable matching becomes student-oriented stable matching.
Therefore, by transforming into a two-sided market with symmetric preferences and finding a match that is stable and has the maximum utility, the student-optimal matching can be found.

In the following sections, we will explain how to solve the stable marriage problem with symmetric preferences.
In section~\ref{sec:reduction}, we will reduce the situation with symmetric preferences to a \textit{assignment problem} by expressing it as a two-dimensional matrix.
In section ~\ref{sec:stability}, we will use the reductio method to prove that the matching generated by reducing it to an assignment problem is a stable matching.
In section~\ref{sec:utility}, we will use the reductiometry method to prove that the preference utility of the generated matching is maximal.
Section~\ref{sec:many-to-one} will explain how to create a many-to-one match.
In section ~\ref{sec:complexity}, we will explain that the proposed method can operate at polynomial time complexity along with pseudocode.

\subsection{Assignment problem reduction} \label{sec:reduction}
To simplify the problem, we assume that each school has one student, and that the number of students and the total number of students in each school are the same.
Since the preference ranks in both directions are the same, the situation in Figure~\ref{fig:sym_inst} can be expressed as a single preference rank matrix like Figure~\ref{fig:rank_matrix}.

\begin{figure}[h]
    \centering
    \begin{tabular}{c c c c}
          & A & B & C \\
        a & 1 & 2 & 3 \\
        b & 1 & 1 & 2 \\
        c & 2 & 1 & 1
    \end{tabular}
    \caption{Preference rank matrix}
    \label{fig:rank_matrix}
\end{figure}

Next, the preference ranking matrix is converted into a weight matrix using the weight conversion function $w$.
At this time, the weight conversion function $w$ must satisfy the following conditions.

\begin{equation} \label{def:weight}
    \begin{split}
        r(s, h) < r(s, h') \land & r(s, h) < r(s', h) \\
        & \implies w(s, h) > w(s', h) + w(s, h') ,\  w \ge 0
    \end{split}
\end{equation}

There are functions such as $w(s, h) = (|\mathbb{S}| + 1)^{z - r(s, h)}$ as weight conversion functions.
Through the weight conversion function, the preference ranking matrix of Figure~\ref{fig:rank_matrix} can be converted to a weight matrix as shown in Figure~\ref{fig:weight_matrix}.

\begin{figure}[h]
    \centering
    \begin{tabular}{c c c c}
          & A & B & C \\
        a & $100_{(4)}$ & $10_{(4)}$ & $1_{(4)}$ \\
        b & $100_{(4)}$ & $100_{(4)}$ & $10_{(4)}$ \\
        c & $10_{(4)}$ & $100_{(4)}$ & $100_{(4)}$
    \end{tabular}
    \caption{Weight matrix, $w(s, h) = (|\mathbb{S}| + 1)^{z - r(s, h)} ,\ |\mathbb{S}| = 3 ,\ z = 3$}
    \label{fig:weight_matrix}
\end{figure}

Finally, we find the matching $M^*$ with the maximum sum of weights.

\begin{equation}
    M^* = \underset{M}{\mathrm{argmax}}\, \sum\limits_{(s, h) \in M}{w(s, h)}
\end{equation}

This problem is known as the allocation problem, minimum cost flow problem, etc.
It is known that in the assignment problem, a perfect match with the maximum sum of weights can be found in polynomial time\cite{kuhn1955hungarian, gabow1989faster, tomizawa1971some, edmonds1972theoretical, crouse2016implementing}.

\subsection{Stability} \label{sec:stability}
\begin{theorem} \label{the:stability}
        Matching $M^*$ is a stable matching.
\end{theorem}
\begin{proof} \label{pro:stability}
    Let assume that the match $M^*$ with the maximum sum of weights is not a weakly stable match.

    By the definition of a stable match, there must be at least one blocking pair in the match $M^*$.
    Let the pair $(s, h)$ included in the matching $M^*$ be a blocking pair, and the pair $(s', h') \in M^*$, then the pair $(s, h'), (s', h) \notin M^*$.
    Then let convert the pair $(s, h'), (s', h)$ of $M^*$ into the pair $(s, h), (s', h')$ which is the matching $M'$.
    \begin{center}
        $M' = M^* - \{(s, h'), (s', h)\} + \{(s, h), (s', h')\}$.
    \end{center}

    By the definition of a blocking pair, student $s$ more prefers school $h$ to school $h'$, and school $h$ more prefers student $s$ to student $s'$:
    \begin{center}
        $r(s, h) < r(s, h')$ and $r(h, s) < r(h, s')$.
    \end{center}

    Since it is a two-sided market with symmetric preferences, we can rewrite it as follows:
    \begin{center}
        $r(s, h) < r(s, h')$ and $r(s, h) < r(s', h)$.
    \end{center}

    The following relationship is established by the weight conversion function definition (Equation~\ref{def:weight}):
    \begin{center}
        $w(s, h') + w(s', h) < w(s, h) + w(s', h')$.
    \end{center}

    Therefore, the following relationship holds between the sum of the weights of the matching $M'$ and the sum of the weights of the matching $M^*$:
    \begin{center}
        $\sum\limits_{M'}{w} = \sum\limits_{M^*}{w} - (w(s, h') + w(s', h)) + (w(s, h) + w(s', h'))$

        $\therefore \sum\limits_{M'}{w} > \sum\limits_{M^*}{w}$.
    \end{center}

    However, since the matching $M^*$ is a match with the maximum total weight, there cannot be a matching $M'$ whose total weight is greater than the matching $M^*$.
    In other words, the assumption that the matching $M^*$ is not stable creates a contradiction in that there exists a matching $M'$ with a larger total weight than the matching $M^*$.
    Therefore, the matching $M^*$ is a stable matching.
\end{proof}

\subsection{Maximum utility} \label{sec:utility}
\begin{theorem} \label{the:utility}
    The matching $M^*$ is the matching where the preference utility is maximized.
\end{theorem}
\begin{proof} \label{pro:utility}
    Assume that the matching $M^*$ with the maximum sum of weights does not maximize the utility.

    Since the utility of the matching $M^*$ is not maximal, there exists a matching $M'$ whose utility is larger:
    \begin{center}
        $Q(M') > Q(M^*)$.
    \end{center}

    At least one pair that is present in the matching $M'$ but not in the matching $M^*$ must have a larger utility.
    Let the pair $(s, h)$ in the match $M'$ has a greater utility than the pair $(s, h')$ and $(s', h)$ which are in the match $M^*$:
    \begin{center}
        $r(s, h) < r(s, h')$ and $r(s, h) < r(s', h)$.
    \end{center}

    By the definition of the weight conversion function (Equation~\ref{def:weight}), the following holds true:
    \begin{center}
        $w(s, h) + w(s', h') > w(s, h') + w(s', h)$.
    \end{center}

    Therefore, the following relationship holds between the sum of the weights of the matching $M'$ and the sum of the weights of the matching $M^*$:
    \begin{center}
        $\sum\limits_{M'}{w} = \sum\limits_{M^*}{w} - (w(s, h') + w(s', h)) + (w(s, h) + w(s', h'))$

        $\therefore \sum\limits_{M'}{w} > \sum\limits_{M^*}{w}$.
    \end{center}

    However, since the matching $M^*$ is a match with the maximum total weight, there cannot be a matching $M'$ whose total weight is greater than the matching $M^*$.
    In other words, the assumption that matching $M^*$ does not maximize the preference utility creates a contradiction in that there exists a matching $M'$ with a larger weight sum than matching $M^*$.
    Therefore, the matching $M^*$ is the matching with the maximum preference utility.
\end{proof}

\subsection{Many-to-one matching} \label{sec:many-to-one}
In order to expand the school capacity to one or more students, \textit{vacant seat} objects are created equal to the number of students in the school.
Only one student can be assigned to each vacant seat, and the student's preference for the vacant seat must be the same as the preference of the school that created the vacant seat.
After creating an student-optimal one-to-one matching between a student and a vacant seat and then assigning the student to the school that created the vacant seat, an student-optimal many-to-one matching that satisfies the quota conditions is created.

\subsection{Complexity} \label{sec:complexity}
Algorithm~\ref{alg:overview} shows the pseudocode of an algorithm that finds student-optimal matching in a two-sided market where one side has preferences and the other side has quota.
As input, it receives a set of students $\mathbb{S}$, a set of schools $\mathbb{H}$, a preference ranking function $r$, a weight transformation function $w$, and a random seed $seed$, and as output returns the student-optimal matching.
Line \ref{alg:init_vacant} creates vacancies according to the number of seats in each school.
Line \ref{alg:init_rank_matrix} creates a preference ranking matrix for empty seats.
Line \ref{alg:convert_weight_matrix} converts the preference ranking matrix for empty seats into a weight matrix.
Line \ref{alg:assignment} finds the match with the maximum total weight in the weight matrix.
When there are multiple matches with the maximum weight sum, one is returned according to $seed$.
Line \ref{alg:convert_many_to_one}-\ref{alg:return} converts one-to-one matching for vacant seats into many-to-one matching for schools and returns.

\begin{algorithm}[h]
    \caption{Algorithm overview} \label{alg:overview}
    \begin{algorithmic}[1]
        \Function{getStudentOptimalMatching}{$\mathbb{S}, \mathbb{H}, r, w, seed$} \label{alg:fn}
        \State $\mathbb{V} \gets createVacantSeatSet(\mathbb{H})$ \label{alg:init_vacant}
        \State $R \gets createPreferenceRankMatrix(\mathbb{S}, \mathbb{V}, r)$ \label{alg:init_rank_matrix}
        \State $W \gets convertToWeightMatrix(R, w)$ \label{alg:convert_weight_matrix}
        \State $M^* \gets getMaximumWeightMatching(W, seed)$ \label{alg:assignment}
        \State $student\_optimal \gets convertToManyToOneMatching(M^*, V)$ \label{alg:convert_many_to_one}
        \State \Return $student\_optimal$ \label{alg:return}
        \EndFunction
    \end{algorithmic}
\end{algorithm}

When the number of students is $n$, the task of creating vacant seats in Line~\ref{alg:init_vacant} is possible in $O(n)$ because the number of vacancies and the number of students are the same.
If the preference ranking function $r$ is implemented in $O(1)$, Line~\ref{alg:init_rank_matrix} is possible in $O(n^2)$.
If the weight conversion function $w$ is implemented in $O(1)$, Line~\ref{alg:convert_weight_matrix} is possible in $O(n^2)$.
The task of finding a match with the maximum sum of weights in Line~\ref{alg:assignment} is possible in $O(n^3)$\cite{tomizawa1971some, edmonds1972theoretical, crouse2016implementing}.
If the task of finding the school that created the vacant seat is implemented in $O(1)$, the task of Line~\ref{alg:convert_many_to_one} is possible in $O(n)$.
Therefore, the time complexity of the entire algorithm is $O(n^3)$, and the problem of finding student-optimal matching in a two-sided market with one-sided preference can be solved in polynomial time.

\section{Experiment}
\subsection{Experimental environment}
All experiments were run on a system equipped with an AMD Ryzen 7 4800H 8-core 16-thread 2.9GHz CPU, two DDR4 3200MHz 16GB memory, and an NVMe HP SSD EX900 500GB.
The algorithm~\ref{alg:overview} was implemented as software using Ubuntu 22.04.3 LTS, Python 3.10.12, and SciPy 1.11.1.
Input data was randomly generated using 10 seeds when the number of students was 100, 1,000, and 10,000.
The number of schools in the experiment was 5, 5, and 50, respectively, and the number of seats in each school was equally distributed.
The student-oriented stable matching created through Tie-Breaking is denoted as Baseline, and the student-optimal matching is denoted as Student-Optimal.
The weight conversion function used was $w(s, h) = 2^{z - r(s, h)} - 1$.
The preference utility is the average value of the number of students according to the preference ranking of the assigned school, and the execution time is the average value of the time taken to execute the corresponding algorithm.
The input data used in the experiment and the allocation results can be found at GitHub\footnote{Dataset GitHub Link: \url{https://github.com/Seongbeom-Park/stable_marriage_one_sided_preference_data}}.

\subsection{100 students \& 5 schools}
\begin{figure}[h]
    \centering
    \begin{subfigure}{0.49\textwidth}
        \includegraphics[width=\textwidth]{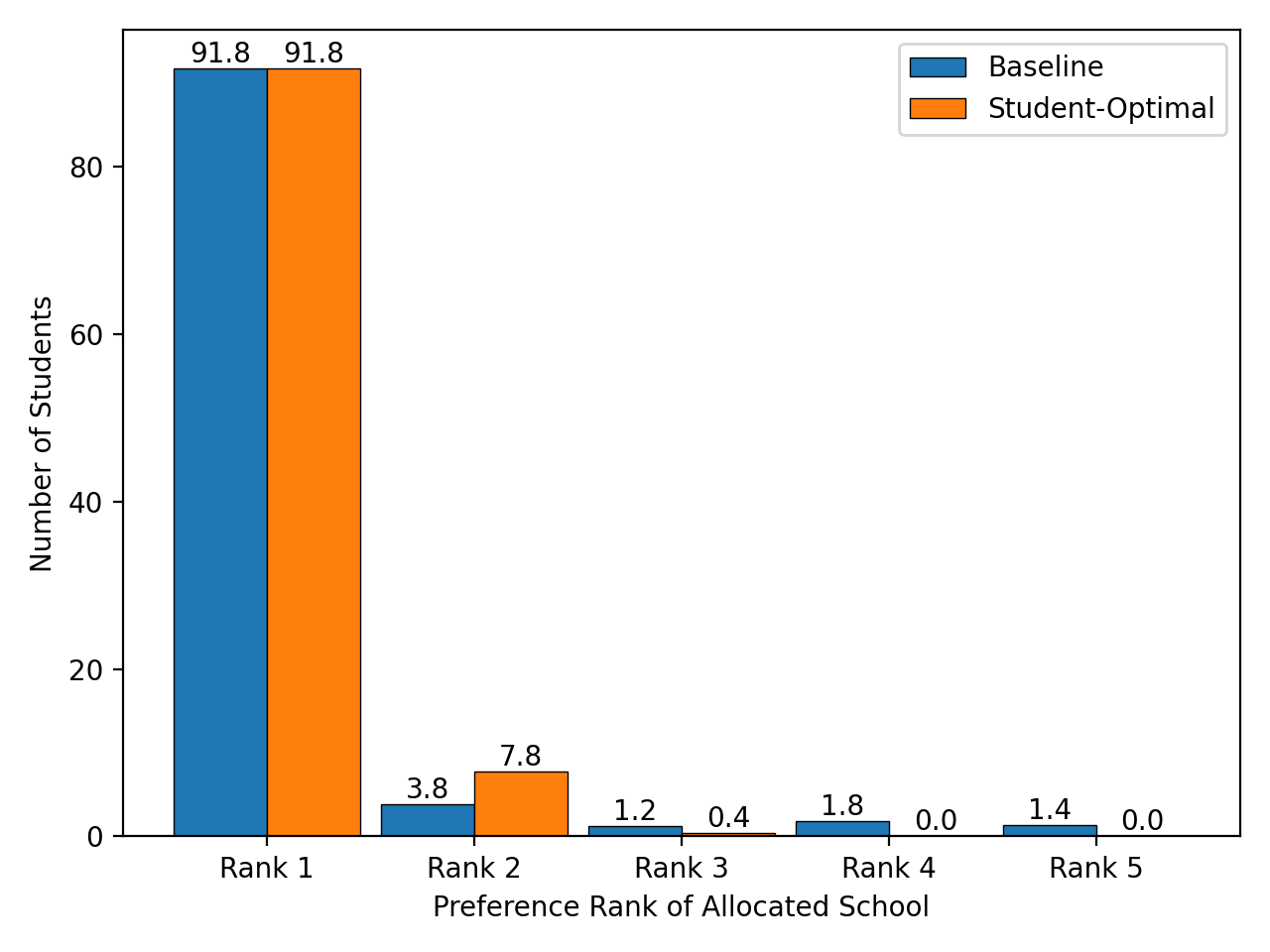}
        \caption{Preference Utility}
        \label{fig:student100_school5_utility}
    \end{subfigure}
    \hfill
    \begin{subfigure}{0.49\textwidth}
        \includegraphics[width=\textwidth]{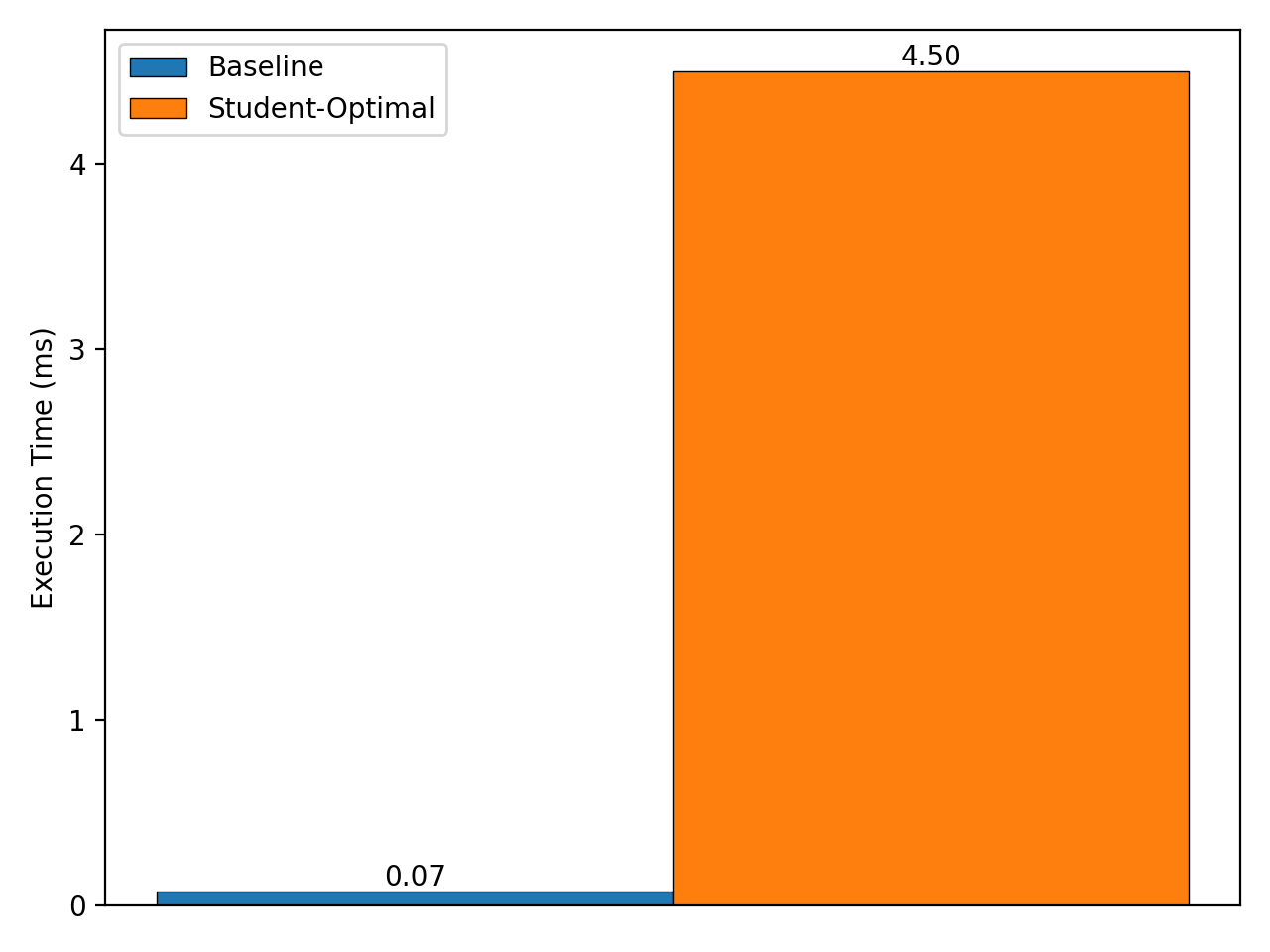}
        \caption{Execution Time}
        \label{fig:student100_school5_exec}
    \end{subfigure}
    \caption{Experiment Results with 100 Students and 5 Schools}
    \label{fig:student100_school5}
\end{figure}

Figure~\ref{fig:student100_school5_utility} shows the preference utility according to assignment method when the number of students is 100 and the number of schools is 5.
The quota for all schools was set at 20 students.
In both Student-Optimal and Baseline, the number of students assigned to the most desired school is the same, with an average of 91.8 students.
However, the number of students assigned to second choice is on average 4 more for Student-Optimal than for Baseline.
In Student-Optimal, 0.4 students are assigned to the 3rd choice, so all students are assigned to a school within the 3rd choice, but in Baseline, 1.2 students are assigned to the 3rd choice, 1.8 students are assigned to the 4th choice, and 1.4 students are assigned to the 5th choice.
The execution time was measured at 4.50ms for Student-Optimal.

\subsection{1,000 students \& 5 schools}
\begin{figure}[h]
    \centering
    \begin{subfigure}{0.49\textwidth}
        \includegraphics[width=\textwidth]{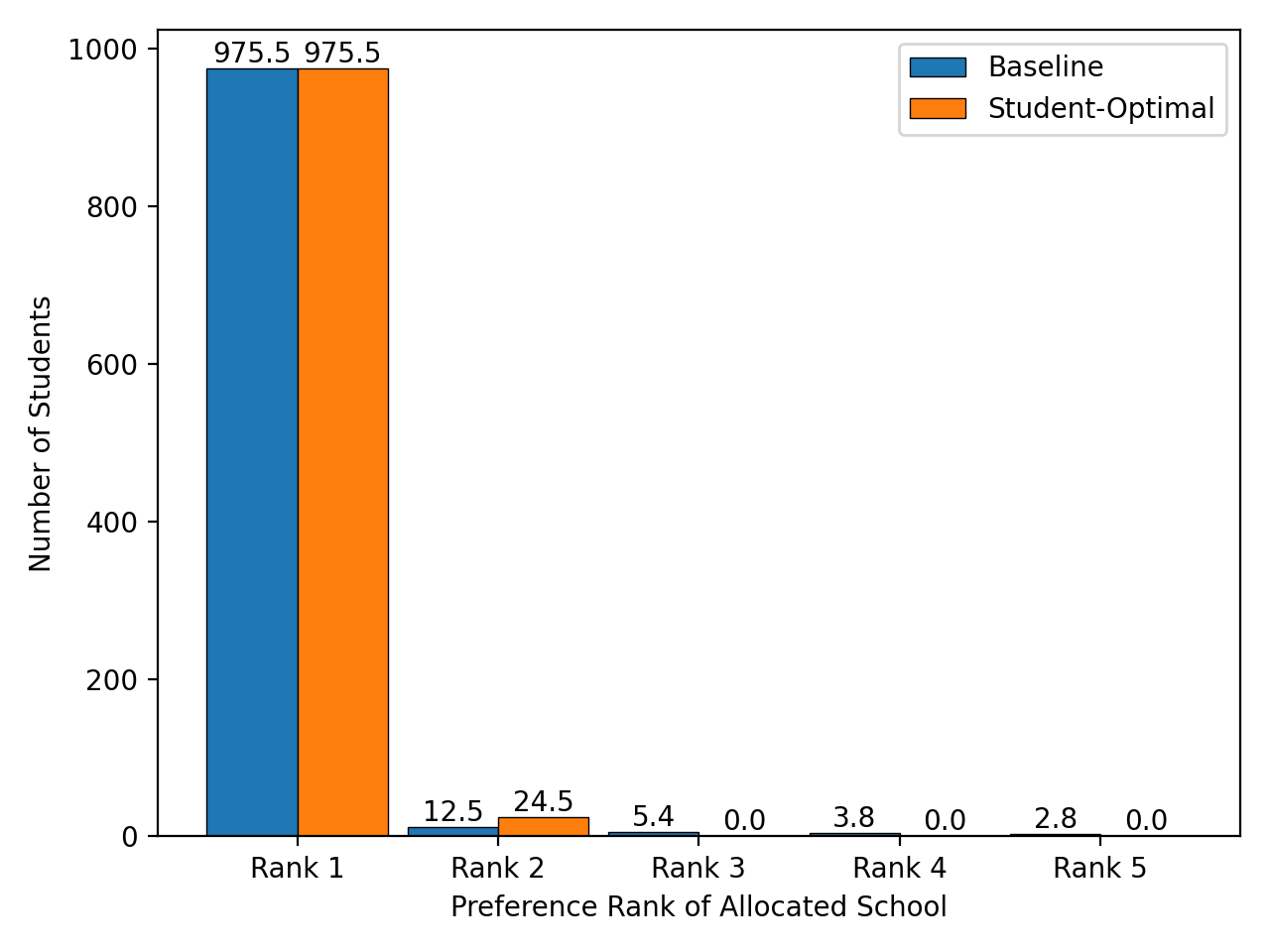}
        \caption{Preference Utility}
        \label{fig:student1000_school5_utility}
    \end{subfigure}
    \hfill
    \begin{subfigure}{0.49\textwidth}
        \includegraphics[width=\textwidth]{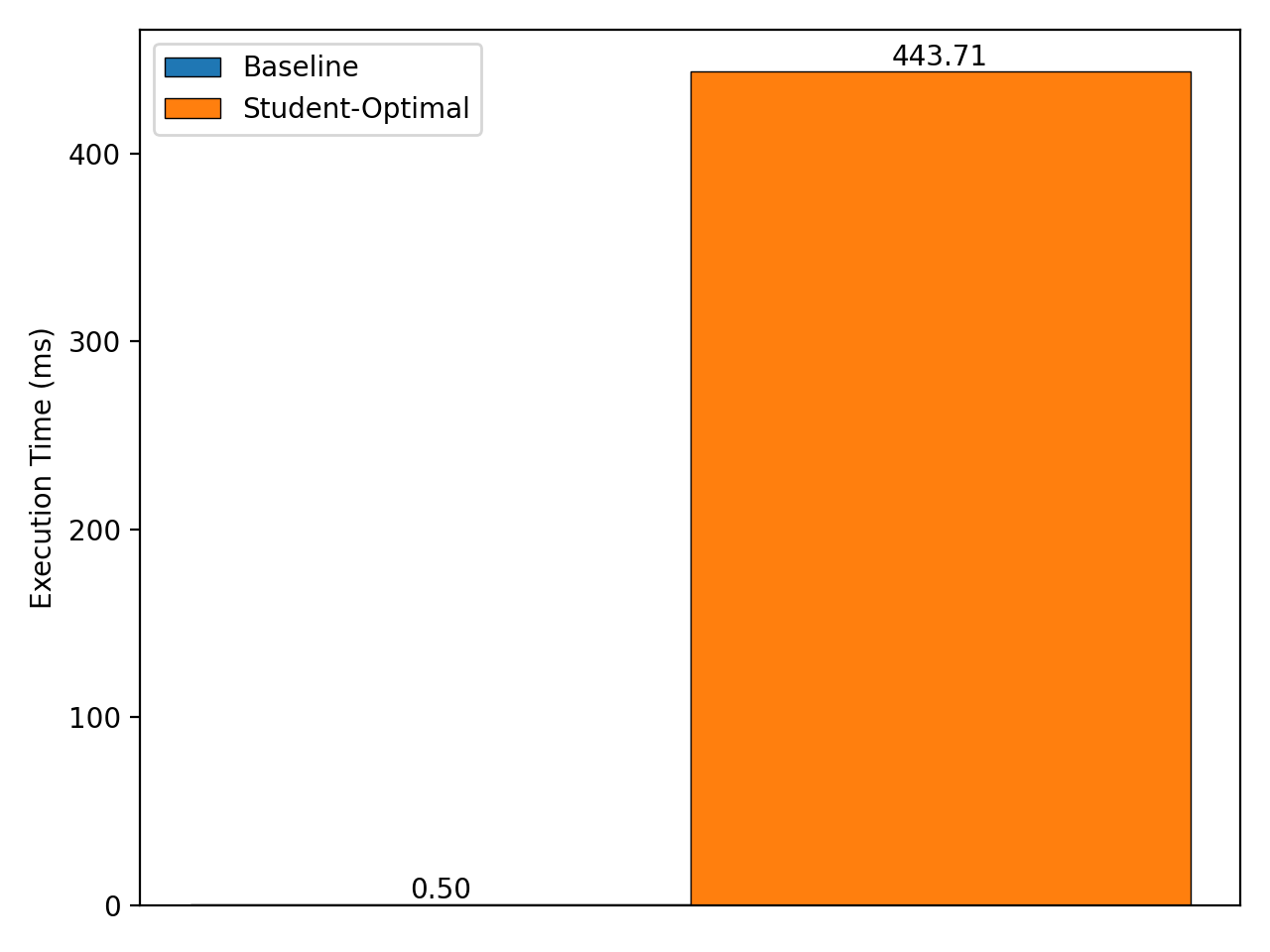}
        \caption{Execution Time}
        \label{fig:student1000_school5_exec}
    \end{subfigure}
    \caption{Experiment Results with 1,000 Students and 5 Schools}
    \label{fig:student1000_school5}
\end{figure}

Figure~\ref{fig:student1000_school5_utility} shows the preference utility according to allocation method when the number of students is 1,000 and the number of schools is 5.
As the number of students increased, the school's capacity also increased to 200.
As the number of students a school can accept increases, the percentage of students assigned to their most desired school increases to 97.55\%.
Student-Optimal assigned an average of 12 more students to second choice than Baseline.
Student-Optimal assigned all students up to the second choice, but Baseline assigned 5.4, 3.8, and 2.8 students to the third, fourth, and fifth choices, respectively.
The execution time was measured to be an average of 443.71ms for Student-Optimal and an average of 0.50ms for Baseline.

\subsection{10,000 students \& 50 schools}
\begin{figure}[h]
    \centering
    \begin{subfigure}{0.49\textwidth}
        \includegraphics[width=\textwidth]{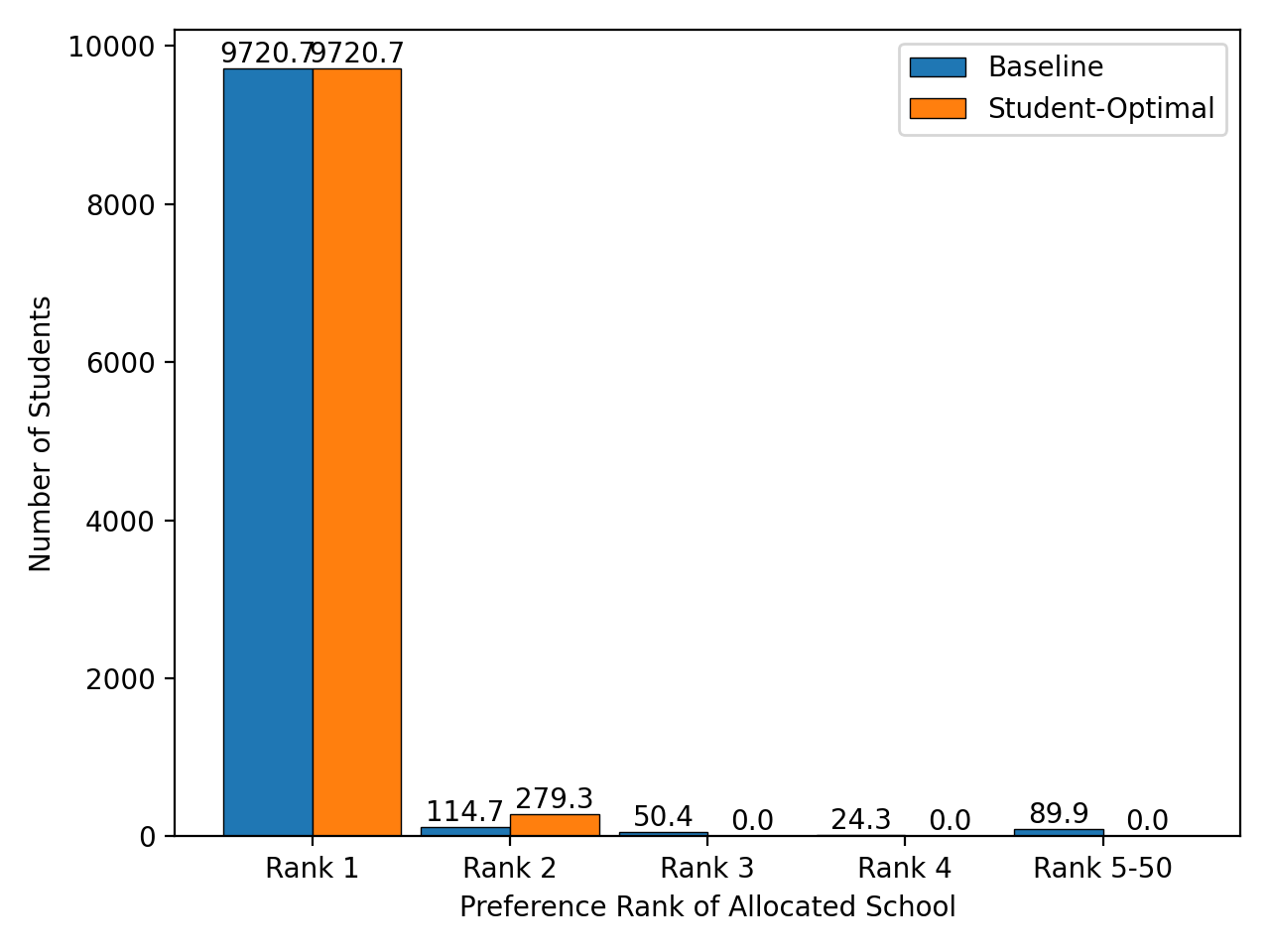}
        \caption{Preference Utility}
        \label{fig:student10000_school50_utility}
    \end{subfigure}
    \hfill
    \begin{subfigure}{0.49\textwidth}
        \includegraphics[width=\textwidth]{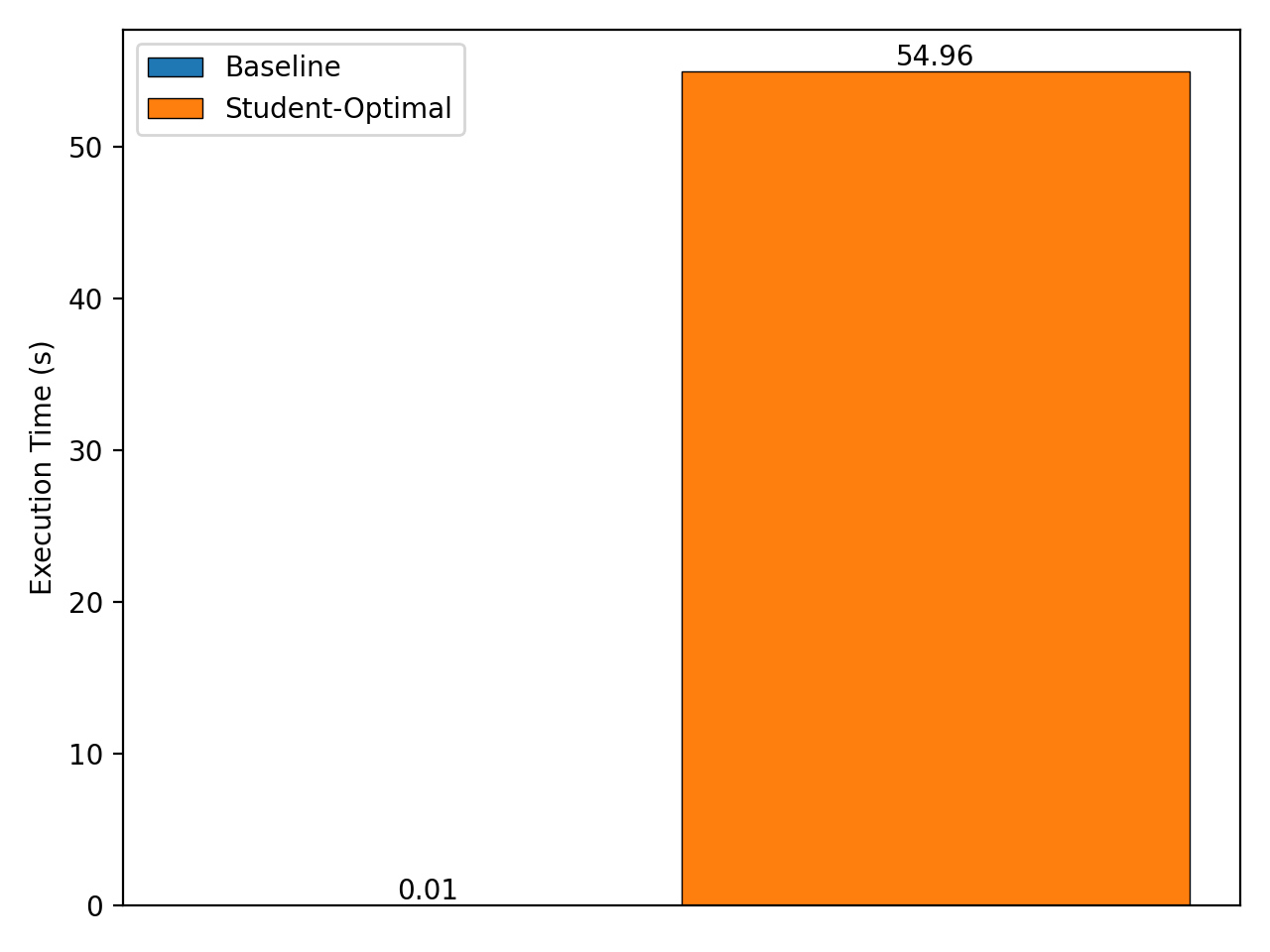}
        \caption{Execution Time}
        \label{fig:student10000_school50_exec}
    \end{subfigure}
    \caption{Experiment Results with 10,000 Students and 50 Schools}
    \label{fig:student10000_school50}
\end{figure}

Figure~\ref{fig:student10000_school50_utility} shows the preference utility according to allocation method when the number of students is 10,000 and the number of schools is 50.
The quota for all schools was set at the same number of 200 students.
An average of 9,720.7 students were assigned to the school they most wanted in both Student-Optimal and Baseline.
Student-Optimal assigned 279.3 students to second choice, and no student was assigned to third choice or lower.
On the other hand, Baseline assigned 114.7, 50.4, and 24.3 people to second, third, and fourth choices, respectively.
Baseline assigned a total of 89.9 people to choices 5-50, with an average of more than 0.1 people being assigned to all choices.
The execution time was measured to be an average of 54.96 seconds for Student-Optimal and 9.58ms for Baseline.

Despite using randomly generated data with no preference or avoidance of specific schools, in several experimental settings, some students were assigned to schools with low baseline scores.
On the other hand, Student-Optimal assigned all students within the third choice when the number of students was 100, and assigned all students within the second choice when the number of students was 1,000 or 10,000.
This is because the student-optimal matching is the match with the maximum preference utility among the student-oriented stable matches that can be created through Tie-Breaking.
Due to relatively high time complexity, the execution time increased significantly as the number of students increased, but if implemented to utilize multi-threads and GPU, the execution time could be shortened.

\section{Conclusion}
Due to the institutional nature of deciding which school to attend without an exam, many ties occur.
Due to the nature of the Gale-Shapley algorithm\cite{gale1962college}, which cannot handle ties, Tie-Breaking\cite{irving1994stable} is conducted through a lottery.
However, when the order of assignment of students is determined randomly, a problem arises in which many students are assigned to schools of low preference\cite{abdulkadiroglu2021school, erdil2008s}.

Because there may be multiple stable matches that can be generated from a given preference, the number of students assigned to each preference rank can vary greatly depending on the outcome of the lottery.
In cases where there is a tie, the problems of finding the best quality match often have a time complexity of NP\cite{abdulkadiroglu2021school, erdil2008s, manlove2002hard, o2007algorithmic}.
Because it is difficult to know how many students' preferences can be met, research has been conducted to improve the quality of the generated matches\cite{erdil2008s, irving2010finding, ravindranath2021deep}.

In this paper, we introduced a method to find student-optimal matching that is student-oriented, stable, and maximizes utility, in a situation where one side has preferences and the other side has limited number of students.
First, it was shown that student-oriented stable matching in a two-sided market with one-sided preferences is the same as the problem of finding stable matching in a two-sided market with symmetric preferences.
In addition, by defining utility from a preference utilitarian perspective\cite{harsanyi1977morality, singer2013companion}, a new method to measure the quality of matching was presented.
Finally, we found the conditions for the weight transformation function to reduce the stable marriage problem with symmetric preference to an assignment problem.

In this paper, it was proven that the matching with the maximum sum of weights is the student-optimal matching.
First, by proving the stability of the generated matching using the reductiometry method, it was proven to be a student-oriented and stable matching.
Additionally, it was proven using the reductio method that the generated matching maximizes the preference utility.
Through this, we proved that in a stable marriage problem where only one side has a preference, a student-oriented, stable, and maximally utility-maximizing student-optimal matching can be found in polynomial time.

In this paper, we quantitatively verified through experiments that the assignment quality was greatly improved.
As the number of students increased from 100 to 10,000, assignment performance was measured using randomly generated preference data.
In the data where many students were assigned to schools with low preference through Tie-Breaking, the students-optimal matching was to assign all students to within their third choice.
In addition, it was verified that it was sufficient for practical use with a performance time of less than 1 minute for 10,000 people.

Lastly, We hope that this research will provide more students with the opportunity to receive the education they want, creating a society where everyone can be a co-winner.

\section*{Acknowledgement}
This research began through discussions about the school assignment system with my father, Changho Park, who has devoted himself to the education field for 33 years.
I also received help on ethics from my younger sister, Seoyoung Park, who teaches students at high school.
And Junghyun Kim, a UNIST alumnus who works at Kakao, also helped develop proofs and mathematical techniques.
The English translation was done with the help of Google Translator.

\medskip

\bibliographystyle{unsrt}
\bibliography{main}

\end{document}


\maketitle

\begin{abstract}
    한국을 포함한 세계 여러 나라에서는 선지원 후추첨제도를 사용하고 있다.
    그러나, 해당 방식은 추첨 결과에 따라 선호도가 떨어지는 학교에 배정되는 학생들이 다수 생기는 문제점이 있다.
    또한, 동점자가 있는 상황에서 좋은 배정을 찾는 작업은 NP의 시간복잡도를 갖는 경우가 많아 배정의 품질을 향상시키기는 매우 어려운 문제다.

    이 연구에서는 한 쪽만 선호도를 갖는 양면 시장에서 학생 중심의 선호 공리를 최대화하는 안정적인 배정을 찾는 문제는 다항시간에 풀 수 있음을 증명하고, 배정의 품질이 향상됨을 실험을 통해 검증한다.
    이 연구의 주요 기여는 다음과 같다.
    한 쪽만 선호도를 갖는 양면 시장에서 학생 중심의 안정적인 배정은 동형 선호도를 갖는 양면 시장의 안정적인 배정과 같음을 발견하였다.
    또한, 선호 공리주의자의 관점에서 배정의 품질을 정량화할 수 있는 방법을 정의하였다.
    위 두 가지를 바탕으로 동형 선호도를 갖는 양면 시장에서 선호 공리를 최대화하는 안정적인 매칭을 찾는 문제가 할당 문제로 환원이 가능함을 증명하였다.
    이 연구에서는 실험을 통해 기존 배정 방식으로는 많은 학생들이 선호도가 떨어지는 학교에 배정되는 상황에서도, 학생 최적의 배정은 더 많은 학생들을 상위 지망의 학교로 배정하는 것을 정량적으로 검증하였다.
\end{abstract}

\section{Introduction}
한국과 미국을 비롯한 세계 여러 나라에서는 시험 없이 학생들을 진학할 학교에 배정할 때, 학생들이 원하는 학교에서 선택할 수 있는 선지원 후추첨 방식이 주로 사용된다.
학생들은 진학을 희망하는 학교들의 목록을 작성하여 제출하고, 지원이 완료되면 배정 담당자가 각 학교의 정원에 맞게 추첨을 통해 학생들을 배정한다.
이 과정에서 모든 학생들을 원하는 학교로 배정하지 못하는데, 원하지 않은 학교에 배정된 학생들은 학습 의욕이 떨어지거나, 통학 거리가 멀어 학교에 적응하기 어려워하는 등 여러 문제점이 발생한다.

일반적으로 학생들을 배정하는 문제는 \textit{안정적인 결혼 문제(stable marriage problem)}로 환원하여 \textit{Gale-Shapley 알고리즘}\cite{gale1962college}을 사용한다.
이는, \textit{양면 시장(two-sided market)}에서 서로에 대한 선호도를 바탕으로 매칭을 생성하는 방법으로, 생성된 매칭의 \textit{안정성(stability)}을 보장한다.
하지만, Gale-Shapley 알고리즘은 선호 목록에서 동점자가 허용되지 않기 때문에, 동점자가 있는 경우에는 \textit{Tie-Breaking}\cite{irving1994stable}을 통해 임의로 동점자를 제거해야 한다.
동점자를 제거하기 위한 방법으로, 무작위 추첨을 통해 학생들에게 임의의 번호를 부여하고, 빠른 번호를 부여받은 학생부터 배정하는 방법이 사용되고 있다.

Gale-Shapley 알고리즘을 사용하면 \textit{안정적인 매칭(stable matching)}이 생성되지만, 추첨 결과에 따라 지망별 배정 인원이 크게 달라진다\cite{abdulkadiroglu2021school}.
왜냐하면 주어진 선호도 목록으로부터 생성 가능한 안정적인 매칭이 여러 개 있을 수 있기 때문이다\cite{irving1987efficient}.
배정의 품질이 얼마나 달라지는지는 다음 예시를 통해 알 수 있다.
Figure~\ref{fig:one_sided_pref}는 학생 3명의 선호 목록을 보여주고 있다.
학생 a는 학교 A, B, C를 순서대로 선호하며, 학생 b, c는 학교 2개를 동시에 첫 번째로 선호하고 나머지 학교를 두 번째로 선호하고 있다.
모든 학교의 정원이 1명인 상황이라고 가정하자.
Figure~\ref{fig:optimal_match}의 매칭 \{(a, A), (a, B), (c, C)\}처럼 배정하면, 학생 3명 전원이 첫 번째로 선호하는 학교로 배정된다.

\begin{figure}[h]
    \begin{subtable}{0.4\textwidth}
        \centering
        \begin{tabular}{c c c c}
            a: & A     & B & C \\
            b: & (A B) & C & ( ) \\
            c: & (B C) & A & ( )
        \end{tabular}
        \caption{Student's preference}
        \label{fig:one_sided_pref}
    \end{subtable}
    \hfill
    \begin{subtable}{0.6\textwidth}
        \centering
        \begin{tabular}{| c || c c c |}
            \hline
            \multirow{2}{*}{Matching} & \multicolumn{3}{c|}{Ranking} \\
                    & 1st & 2nd & 3rd \\
            \hline
            \hline
            \{(a, A), (b, B), (c, C)\} & 3 & 0 & 0 \\
            \hline
        \end{tabular}
        \caption{Student-optimal matching}
        \label{fig:optimal_match}
    \end{subtable}
    \caption{Student-optimal matching with student preference only}
\end{figure}

만약 지원 단계에서 엄격한 선호도(strictly ordered preference)를 요구한다면, 학생 b와 c는 동일하게 선호하는 두 학교를 분리하여 지원해야 한다.
Figure~\ref{fig:strict_student_pref}처럼 학생들이 동점자를 제거한 선호도로 지원하고, Figure~\ref{fig:strict_school_pref}처럼 학교는 모든 학생을 동일하게 선호하는 상황에서, 학생 a와 b의 배정 순서에 따라 \textit{학생 중심(student-oriented)}의 안정적인 매칭은 2가지가 있다.
Figure~\ref{fig:strict_alloc}는 매칭에 따라 지망별로 배정된 학생 수를 보여준다.
동점자가 제거된 조건에서 생성되는 매칭 중에는 Figure~\ref{fig:optimal_match}의 이상적인 매칭이 포함되지 않는다.

\begin{figure}[h]
    \centering
    \begin{subtable}{0.4\textwidth}
        \centering
        \begin{tabular}{c c c c}
            a: & A & B & C \\
            b: & A & B & C \\
            c: & B & C & A
        \end{tabular}
        \caption{Student's preferences}
        \label{fig:strict_student_pref}
    \end{subtable}
    \hfill
    \begin{subtable}{0.4\textwidth}
        \centering
        \begin{tabular}{c c}
            A: & (a b c) \\
            B: & (a b c) \\
            C: & (a b c)
        \end{tabular}
        \caption{School's preferences}
        \label{fig:strict_school_pref}
    \end{subtable}
    \hfill
    \begin{subtable}{\textwidth}
        \centering
        \begin{tabular}{| c || c c c |}
            \hline
            \multirow{2}{*}{Matching} & \multicolumn{3}{c|}{Ranking} \\
            & 1st & 2nd & 3rd \\
            \hline
            \hline
            \{(a, A), (b, C), (c, B)\} & 2 & 0 & 1 \\
            \{(a, C), (b, A), (c, B)\} & 2 & 0 & 1 \\
            \hline
        \end{tabular}
        \caption{Possible student-oriented matchings}
        \label{fig:strict_alloc}
    \end{subtable}
    \caption{Possible matchings with student's strictly ordered preference}
\end{figure}

Figure~\ref{fig:tie_student_pref}처럼 지망별로 복수의 학교들을 지원하고, Figure~\ref{fig:tie_school_pref}처럼 학교는 학생을 동일하게 선호하는 상황에서, 학생들의 배정 순서에 따라 생성되는 안정적인 매칭은 4가지가 있다.
Figure~\ref{fig:tie_alloc}는 매칭에 따라 지망별 배정된 학생 수를 보여준다.
어떤 학생부터 배정하느냐에 따라 모든 학생이 원하는 학교에 배정될 수도 있고, 선호도가 낮은 학교에 배정되는 학생들이 생기기도 한다.

\begin{figure}[h]
    \centering
    \begin{subtable}{0.3\textwidth}
        \centering
        \begin{tabular}{c c c c}
            a: & A     & B & C   \\
            b: & (A B) & C & ( ) \\
            c: & (B C) & A & ( )
        \end{tabular}
        \caption{Student's preferences}
        \label{fig:tie_student_pref}
    \end{subtable}
    \hfill
    \begin{subtable}{0.3\textwidth}
        \centering
        \begin{tabular}{c c}
            A: & (a b c) \\
            B: & (a b c) \\
            C: & (a b c)
        \end{tabular}
        \caption{School's preferences}
        \label{fig:tie_school_pref}
    \end{subtable}
    \hfill
    \begin{subtable}{\textwidth}
        \centering
        \begin{tabular}{| c || c c c |}
            \hline
            \multirow{2}{*}{Matching} & \multicolumn{3}{c|}{Ranking} \\
            & 1st & 2nd & 3rd \\
            \hline
            \hline
            \{(a, A), (b, B), (c, C)\} & 3 & 0 & 0 \\
            \{(a, A), (b, C), (c, B)\} & 2 & 1 & 0 \\
            \{(a, B), (b, A), (c, C)\} & 2 & 1 & 0 \\
            \{(a, C), (b, A), (c, B)\} & 2 & 0 & 1 \\
            \hline
        \end{tabular}
        \caption{Possible student-oriented matchings}
        \label{fig:tie_alloc}
    \end{subtable}
    \caption{Possible matchings with tie-breaking}
\end{figure}

예시의 상황들에서 이상적인 매칭을 생성하기 어려운 이유는 동점자를 제거하는 과정에서 제약조건이 추가되기 때문이다\cite{erdil2008s}.
Tie-Breaking은 배정의 결과를 고려하지 않기 때문에, 무작위 추첨 결과에 따라 추가되는 제약조건이 배정의 품질에 큰 영향을 준다.
최상의 품질을 갖는 배정 방법을 찾기 위해 \textit{평등주의자(egalitarian)}와 \textit{최소 후회(minimum regret)} 등의 관점에서 연구되었지만,
동점자가 있는 경우 시간 복잡도가 비결정론적 다항시간(Non-deterministic Polynomial time, NP)인 경우가 많아\cite{abdulkadiroglu2021school, manlove2002hard, o2007algorithmic},
배정의 품질을 개선시키는 방법들이 연구되고 있다\cite{erdil2008s, irving2010finding, ravindranath2021deep}.

이 연구에서는 한 쪽은 선호도를 갖고 다른 쪽은 정원 제한만 있는 양면 시장에서 학생 중심의 안정적인 매칭 중 공리를 최대화하는 \textit{학생 최적의 매칭}을 다항시간에 찾을 수 있음을 증명할 것이다.
그리고 실험을 통해 배정 품질이 크게 향상됨을 검증할 것이다.
이 연구의 주요 기여는 다음과 같다.
(1) 한 쪽만 선호도를 갖는 양면 시장에서 학생 중심의 안정적인 매칭을 찾는 문제는 동형 선호도를 갖는 양면 시장에서 안정적인 매칭을 찾는 문제와 같음을 보인다.
(2) 선호 공리주의자의 관점에서 공리를 정의하여 매칭의 품질을 정량화한다.
(3) 다항시간에 선호 공리를 최대화하는 안정적인 매칭을 찾는 방법을 제안하고, 제안한 방법으로 찾은 매칭이 학생 최적의 매칭임을 증명한다.
(4) 실험을 통해 학생 최적의 매칭이 배정 품질을 얼마나 향상시키는지 보인다.

\section{The Assignment Criteria}
\subsection{Stable matching}
안정적인 매칭은 \textit{깨트리는 쌍(blocking pair)}이 없는 매칭으로 정의된다\cite{gale1962college}.
깨트리는 쌍이란, 매칭에 존재하지 않는 어떤 쌍이 현재 파트너보다 서로를 더 선호하여 기존 쌍들을 깨트리는 쌍을 의미한다.
동점자가 있는 양면 시장에서는 같은 정도로 선호하는 쌍도 깨트리는 쌍으로 볼 것인지에 따라 안정성이 세 종류(weakly, strongly, super)로 확장된다\cite{irving1994stable}.
강한/슈퍼 안정적인 매칭에서는 같은 정도로 선호하는 경우에도 깨트리는 쌍으로 보지만, \textit{약하게 안정적인 매칭(weakly stable matching)}에서는 더 선호하는 경우에만 깨트리는 쌍으로 본다.
이 연구에서는 모든 학생들을 배정할 수 있는 약하게 안정적인 매칭을 찾는 것을 목표로 한다.

\subsection{Student-oriented}
Gale-Shapley 알고리즘은 지원하는 쪽에게 유리한 매칭을 생성한다\cite{irving1987efficient, abdulkadirouglu2005new}.
학생에게 유리한 배정을 생성하기 위해 학생들이 더 선호하는 학교부터 순서대로 지원하고,
학교는 모든 학생들을 동일하게 선호하므로 빠른 배정순서를 받은 학생부터 학교에 배정하여 정원이 모두 찰 때까지 승인한다.
결과적으로, 학생 중심적인 매칭은 낮은 순위로 지원한 학생보다 높은 순위로 지원한 학생을 우선적으로 학교에 배정한다.

어떤 학생이 다른 학생보다 우선적으로 배정되어야 하는 상황은 학교의 선호도로 표현할 수 있다.
Figure~\ref{fig:sym_inst}는 학생의 학교에 대한 선호도와, 학교의 학생에 대한 선호도가 일치하는 양면 시장을 보여주고 있다.
학교의 선호도를 통해 학생 중심적인 매칭이 생성되도록 재구성하였으므로, 이러한 선호도에서 생성되는 안정적인 매칭은 항상 학생 중심적이다.

\begin{figure}[h]
    \centering
    \begin{subtable}{0.3\textwidth}
        \centering
        \begin{tabular}{c c c c}
            a: & A     & B & C   \\
            b: & (A B) & C & ( ) \\
            c: & (B C) & A & ( )
        \end{tabular}
        \caption{Student's preferences}
        \label{fig:sym_student_pref}
    \end{subtable}
    \hfill
    \begin{subtable}{0.3\textwidth}
        \centering
        \begin{tabular}{c c c c}
            A: & (a b) & c & ( ) \\
            B: & (b c) & a & ( ) \\
            C: & c & b & a
        \end{tabular}
        \caption{School's preferences}
        \label{fig:sym_school_pref}
    \end{subtable}
    \caption{An instance with symmetric preferences}
    \label{fig:sym_inst}
\end{figure}

Figure~\ref{fig:sym_inst}에서 $i$가 $j$를 선호하는 순위를 나타내는 함수를 $r(i, j)$라고 할 때, 학생 $s$와 학교 $h$에서 $r(s, h) = r(h, s)$가 성립한다.
예를 들어, 학생 a가 학교 A를 선호하는 순위($r(\text{a}, \text{A}) = 1$)는 학교 A가 학생 a를 선호하는 순위($r(\text{A}, \text{a}) = 1$)와 같다.
이처럼 서로의 선호도가 일치하는 경우 동형 선호도라고 한다\cite{gai2007acyclic}.

\subsection{Maximizing utility}
John Harsanyi는 \textit{선호 공리주의(preference utilitarianism)}란 \textit{선호 자율성(preference autonomy)}이라는 철학적 원칙을 공리주의로 구성한 것이라 하였다\cite{harsanyi1977morality}.
선호 자율성이라는 원칙하에 좋고 나쁨을 결정하는 것은 오직 개인의 선호도에 달려있다.
모든 학생들을 만족시킬 수 있다면 가장 좋겠지만, 어떤 학교에 정원보다 많은 학생들이 지원하면 일부 학생들은 떨어질 수 밖에 없다.
사회 구성원 간 선호가 상충하는 경우, 개인의 선호도에 일정한 가중치를 할당하여 가장 높은 수준의 만족도를 제공하는 결과를 찾는 방식으로 선호 공리주의를 적용할 수 있다\cite{singer2013companion}.
이 연구에서는 배정된 학교의 선호 순위가 학생의 만족도에 큰 영향을 미친다고 보아, 선호 순위에 따른 가중치의 차이를 크게 설정하였다.

매칭 $M$의 선호 공리 $Q$는 학생 집합 $\mathbb{S}$에 대해 $(|\mathbb{S}| + 1)$-진수로 정의한다.
공리 $Q(M)$의 각 자리수 $q_k$는 매칭 $M$에서 배정된 파트너의 선호 순위가 $k$인 쌍의 개수를 의미하며, $z$는 학생의 선호 목록 길이의 최대값을 의미한다.

\begin{center}
    $Q(M) = \sum\limits_{k=1}^{z}{q_k (|\mathbb{S}| + 1)^{z - k}}$
\end{center}

\begin{center}
    $q_k = \lvert \{(s, h) \in M \mid r(s, h) = k\} \rvert ,\  0 \le q_k \le |\mathbb{S}|$
\end{center}

\begin{center}
    $z = \max\limits_{s \in \mathbb{S}, h \in \mathbb{H}}{r(s, h)}$
\end{center}

예를 들어, Figure~\ref{fig:tie_alloc}에서 $Q(\text{\{(a, A), (b, B), (c, C)\}}) = 300_{(4)}$으로 공리가 가장 크고,
$Q(\text{\{(a, A), (b, C), (c, B)\}}) = Q(\text{\{(a, B), (b, A), (c, C)\}}) = 210_{(4)}$으로 공리가 같으며,
$Q(\text{\{(a, C), (b, A), (c, B)\}}) = 201_{(4)}$로 공리가 가장 작다.

동형 선호도를 갖는 양면 시장에서 평등주의 또는 최소 후회를 최대한 만족시키는 안정적인 매칭을 찾는 문제의 시간 복잡도는 NP-hard로 알려져 있다\cite{o2007algorithmic}.
이 연구에서는 동형 선호도를 갖는 양면 시장에서 선호 공리를 최대화하는 안정적인 매칭을 찾는다.

\section{Stabale Marriage and Assigment Problem}
이 섹션에서는 학생 집합 $\mathbb{S}$만 선호도를 가지고, 학교 집합 $\mathbb{H}$는 정원 제한만 있는 양면 시장에서 학생 최적의 매칭을 찾을 것이다.
학생 최적의 매칭은 학생 중심의 안정적인 매칭 중 공리가 최대인 매칭을 의미한다.
한 쪽만 선호도가 있는 상황은 동형 선호도를 가진 상황으로 변형하면 안정적인 매칭이 학생 중심적인 안정적이 매칭이 된다.
따라서, 동형 선호도를 갖는 양면 시장으로 변형하여 안정적인 동시에 공리가 최대인 매칭을 찾으면, 학생 최적의 매칭을 찾을 수 있다.

이어지는 섹션에서는 동형 선호도를 갖는 안정적인 결혼 문제를 푸는 방법에 대해서 설명할 것이다.
섹션~\ref{sec:reduction}에서는 동형 선호도를 갖는 상황을 2차원 행렬로 표현함으로써 \textit{할당 문제(assignment problem)}로 환원할 것이다.
섹션~\ref{sec:stability}에서는 할당 문제로 환원하여 생성한 매칭이 안정적인 매칭임을 귀류법을 사용하여 증명할 것이다.
섹션~\ref{sec:utility}에서는 생성된 매칭의 선호 공리가 최대임을 귀류법을 사용하여 증명할 것이다.
섹션~\ref{sec:many-to-one}에서는 다대일 매칭을 생성하는 방법에 대해서 설명할 것이다.
섹션~\ref{sec:complexity}에서는 제안한 방법의 의사코드와 함께 다항시간의 시간복잡도에 동작이 가능함을 설명할 것이다.

\subsection{Assignment problem reduction} \label{sec:reduction}
문제를 단순화하기 위해 모든 학교의 정원이 1명이며, 학생 수와 학교의 정원 총합이 같은 상황을 가정한다.
양 방향의 선호 순위가 동일하므로, Figure~\ref{fig:sym_inst}의 상황은 Figure~\ref{fig:rank_matrix}처럼 하나의 선호 순위 행렬로 표현할 수 있다.

\begin{figure}[h]
    \centering
    \begin{tabular}{c c c c}
          & A & B & C \\
        a & 1 & 2 & 3 \\
        b & 1 & 1 & 2 \\
        c & 2 & 1 & 1
    \end{tabular}
    \caption{Preference rank matrix}
    \label{fig:rank_matrix}
\end{figure}

다음으로 선호 순위 행렬을 가중치 변환 함수 $w$를 사용하여 가중치 행렬로 변환한다.
이때, 가중치 변환 함수 $w$는 다음 조건을 만족해야 한다.

\begin{equation} \label{def:weight}
    \begin{split}
        r(s, h) < r(s, h') \land & r(s, h) < r(s', h) \\
        & \implies w(s, h) > w(s', h) + w(s, h') ,\  w \ge 0
    \end{split}
\end{equation}

가중치 변환 함수로 $w(s, h) = (|\mathbb{S}| + 1)^{z - r(s, h)}$와 같은 함수들이 있다.
가중치 변환 함수를 통해 Figure~\ref{fig:rank_matrix}의 선호 순위 행렬은 Figure~\ref{fig:weight_matrix}과 같이 가중치 행렬로 변환이 가능하다.

\begin{figure}[h]
    \centering
    \begin{tabular}{c c c c}
          & A & B & C \\
        a & $100_{(4)}$ & $10_{(4)}$ & $1_{(4)}$ \\
        b & $100_{(4)}$ & $100_{(4)}$ & $10_{(4)}$ \\
        c & $10_{(4)}$ & $100_{(4)}$ & $100_{(4)}$
    \end{tabular}
    \caption{Weight matrix, $w(s, h) = (|\mathbb{S}| + 1)^{z - r(s, h)} ,\ |\mathbb{S}| = 3 ,\ z = 3$}
    \label{fig:weight_matrix}
\end{figure}

마지막으로, 가중치의 총합이 최대인 매칭 $M^*$를 찾는다.

\begin{equation}
    M^* = \underset{M}{\mathrm{argmax}}\, \sum\limits_{(s, h) \in M}{w(s, h)}
\end{equation}

이 문제는 할당 문제, 최소비용 흐름 문제 등으로 알려진 문제다.
할당 문제에서 가중치의 총합이 최대인 완벽한 매칭은 다항시간에 찾을 수 있음이 알려져 있다\cite{kuhn1955hungarian, gabow1989faster, tomizawa1971some, edmonds1972theoretical, crouse2016implementing}.

\subsection{Stability} \label{sec:stability}
\begin{theorem} \label{the:stability}
    매칭 $M^*$는 안정적인 매칭이다.
\end{theorem}
\begin{proof} \label{pro:stability}
    가중치의 총합이 최대인 매칭 $M^*$가 약하게 안정적인 매칭이 아니라고 가정하자.

    안정적인 매칭의 정의에 의해 매칭 $M^*$에는 깨트리는 쌍이 적어도 하나 존재해야 한다.
    매칭 $M^*$에 포함된 쌍 $(s, h)$가 깨트리는 쌍이라 하고, 쌍 $(s', h') \in M^*$라 하면, 쌍 $(s, h'), (s', h) \notin M^*$이다.
    그러면 우리는 $M^*$의 쌍 $(s, h'), (s', h)$를, 쌍 $(s, h), (s', h')$으로 변환한 매칭 $M'$을 만들 수 있다.
    \begin{center}
        $M' = M^* - \{(s, h'), (s', h)\} + \{(s, h), (s', h')\}$.
    \end{center}

    깨트리는 쌍의 정의에 의해 학생 $s$는 학교 $h'$보다 학교 $h$를 더 선호하고, 학교 $h$는 학생 $s'$보다 학생 $s$를 더 선호한다:
    \begin{center}
        $r(s, h) < r(s, h')$ and $r(h, s) < r(h, s')$.
    \end{center}

    동형 선호도를 갖는 양면시장이므로, 다음과 같이 바꿔쓸 수 있다:
    \begin{center}
        $r(s, h) < r(s, h')$ and $r(s, h) < r(s', h)$.
    \end{center}

    가중치 변환 함수 정의(수식~\ref{def:weight})에 의해 다음 관계가 성립한다:
    \begin{center}
        $w(s, h') + w(s', h) < w(s, h) + w(s', h')$.
    \end{center}

    따라서, 매칭 $M'$의 가중치 총합과 매칭 $M^*$의 가중치 총합 사이에는 다음 관계가 성립한다:
    \begin{center}
        $\sum\limits_{M'}{w} = \sum\limits_{M^*}{w} - (w(s, h') + w(s', h)) + (w(s, h) + w(s', h'))$

        $\therefore \sum\limits_{M'}{w} > \sum\limits_{M^*}{w}$.
    \end{center}

    하지만, 매칭 $M^*$는 가중치의 총합이 최대인 매칭이므로, 매칭 $M^*$보다 가중치의 총합이 더 큰 매칭 $M'$은 존재할 수 없다.
    즉, 매칭 $M^*$가 안정적이지 않다는 가정은 매칭 $M^*$보다 가중치 총합이 더 큰 매칭 $M'$이 존재한다는 모순을 만든다.
    따라서, 매칭 $M^*$는 안정적인 매칭이다.
\end{proof}

\subsection{Maximum utility} \label{sec:utility}
\begin{theorem} \label{the:utility}
    매칭 $M^*$는 선호 공리가 최대인 매칭이다.
\end{theorem}
\begin{proof} \label{pro:utility}
    가중치의 총합이 최대인 매칭 $M^*$가 공리를 최대화하지 않는다고 가정하자.

    매칭 $M^*$의 공리가 최대가 아니므로, 공리가 더 큰 매칭 $M'$이 존재한다:
    \begin{center}
        $Q(M') > Q(M^*)$.
    \end{center}

    매칭 $M^*$에는 없고 매칭 $M'$에만 있는 적어도 한 쌍이 더 큰 공리를 가져야 한다.
    매칭 $M'$에만 있는 쌍 $(s, h)$가, 매칭 $M^*$에 있는 쌍 $(s, h')$과 쌍 $(s', h)$보다 큰 공리를 갖는다고 하자:
    \begin{center}
        $r(s, h) < r(s, h')$ and $r(s, h) < r(s', h)$.
    \end{center}

    가중치 변환 함수의 정의(수식~\ref{def:weight})에 의해 다음이 성립한다:
    \begin{center}
        $w(s, h) + w(s', h') > w(s, h') + w(s', h)$.
    \end{center}

    따라서, 매칭 $M'$의 가중치 총합과 매칭 $M^*$의 가중치 총합 사이에는 다음 관계가 성립한다:
    \begin{center}
        $\sum\limits_{M'}{w} = \sum\limits_{M^*}{w} - (w(s, h') + w(s', h)) + (w(s, h) + w(s', h'))$

        $\therefore \sum\limits_{M'}{w} > \sum\limits_{M^*}{w}$.
    \end{center}

    하지만, 매칭 $M^*$는 가중치의 총합이 최대인 매칭이므로, 매칭 $M^*$보다 가중치의 총합이 더 큰 매칭 $M'$은 존재할 수 없다.
    즉, 매칭 $M^*$가 선호 공리를 최대화하지 않는다는 가정은 매칭 $M^*$보다 가중치 총합이 더 큰 매칭 $M'$이 존재한다는 모순을 만든다.
    따라서, 매칭 $M^*$는 선호 공리가 최대인 매칭이다.
\end{proof}

\subsection{Many-to-one matching} \label{sec:many-to-one}
학교 정원이 1명 이상인 경우로 확장시키기 위해, 학교의 정원 수만큼 \textit{빈자리} 객체를 생성한다.
각 빈자리에는 한 명의 학생만 배정될 수 있으며, 학생의 빈자리에 대한 선호도는 빈자리를 생성한 학교의 선호도와 같도록 한다.
학생과 빈자리에 대한 학생 최적의 일대일 매칭을 생성한 뒤, 해당 빈자리를 생성한 학교로 학생을 배정하면 정원 조건을 만족하는 학생 최적의 다대일 매칭이 생성된다.

\subsection{Complexity} \label{sec:complexity}
알고리즘~\ref{alg:overview}은 한 쪽은 선호도를 가지고 다른 쪽은 정원을 가지는 양면 시장에서 학생 최적의 매칭을 찾는 알고리즘의 의사코드를 보여주고 있다.
입력으로 학생 집합 $\mathbb{S}$, 학교 집합 $\mathbb{H}$, 선호 순위 함수 $r$, 가중치 변환 함수 $w$, 랜덤 시드 $seed$를 받고, 출력으로 학생 최적의 매칭을 반환한다.
Line \ref{alg:init_vacant}에서는 각 학교의 정원에 따라 빈자리들을 생성한다.
Line \ref{alg:init_rank_matrix}에서는 빈자리에 대한 선호 순위 행렬을 생성한다.
Line \ref{alg:convert_weight_matrix}에서는 빈자리에 대한 선호 순위 행렬을 가중치 행렬로 변환한다.
Line \ref{alg:assignment}에서는 가중치 행렬에서 가중치의 총합이 최대인 매칭을 찾는다.
가중치 합이 최대인 매칭이 여러 개 있을 때 $seed$에 따라 하나를 반환한다.
Line \ref{alg:convert_many_to_one}-\ref{alg:return}에서는 빈자리에 대한 일대일 매칭을 학교에 대한 다대일 매칭으로 변환하여 반환한다.

\begin{algorithm}[h]
    \caption{Algorithm overview} \label{alg:overview}
    \begin{algorithmic}[1]
        \Function{getStudentOptimalMatching}{$\mathbb{S}, \mathbb{H}, r, w, seed$} \label{alg:fn}
        \State $\mathbb{V} \gets createVacantSeatSet(\mathbb{H})$ \label{alg:init_vacant}
        \State $R \gets createPreferenceRankMatrix(\mathbb{S}, \mathbb{V}, r)$ \label{alg:init_rank_matrix}
        \State $W \gets convertToWeightMatrix(R, w)$ \label{alg:convert_weight_matrix}
        \State $M^* \gets getMaximumWeightMatching(W, seed)$ \label{alg:assignment}
        \State $student\_optimal \gets convertToManyToOneMatching(M^*, V)$ \label{alg:convert_many_to_one}
        \State \Return $student\_optimal$ \label{alg:return}
        \EndFunction
    \end{algorithmic}
\end{algorithm}

학생 수를 $n$이라 할 때, Line~\ref{alg:init_vacant}에서 빈자리를 생성하는 작업은 빈자리 개수와 학생 수가 같으므로 $O(n)$에 가능하다.
선호 순위 함수 $r$을 $O(1)$로 구현하면, Line~\ref{alg:init_rank_matrix}은 $O(n^2)$에 가능하다.
가중치 변환 함수 $w$를 $O(1)$로 구현하면, Line~\ref{alg:convert_weight_matrix}는 $O(n^2)$에 가능하다.
Line~\ref{alg:assignment}에서 가중치 총합이 최대인 매칭을 찾는 작업은 $O(n^3)$에 가능하다\cite{tomizawa1971some, edmonds1972theoretical, crouse2016implementing}.
빈자리를 생성한 학교를 찾는 작업을 $O(1)$로 구현하면, Line~\ref{alg:convert_many_to_one}의 작업은 $O(n)$에 가능하다.
따라서, 전체 알고리즘의 시간 복잡도가 $O(n^3)$으로, 한 쪽만 선호도를 갖는 양면 시장에서 학생 최적의 매칭을 찾는 문제는 다항시간에 해결이 가능하다.

\section{Experiment}
\subsection{Experimental environment}
모든 실험은 AMD Ryzen 7 4800H 8core 16thread 2.9GHz CPU, 2개의 DDR4 3200MHz 16GB 메모리, NVMe HP SSD EX900 500GB를 탑재한 시스템에서 실행하였다.
Ubuntu 22.04.3 LTS, Python 3.10.12, SciPy 1.11.1을 사용하여 알고리즘~\ref{alg:overview}을 소프트웨어로 구현하였다.
입력 데이터는 학생 수 100명, 1,000명, 10,000명일 때, 10개의 시드를 사용하여, 무작위로 학생의 선호도 데이터를 생성하였다.
학교 수는 각각 5개, 5개, 50개로 실험하였으며, 각 학교의 정원은 균등하게 분배하였다.
Tie-Breaking을 통해 생성된 학생 중심의 안정적인 매칭은 Baseline으로, 학생 최적의 매칭은 Student-Optimal로 표기하였다.
가중치 변환 함수는 $w(s, h) = 2^{z - r(s, h)} - 1$을 사용하였다.
선호 공리는 배정된 학교의 선호 순위에 따른 학생 수의 평균값이며, 수행시간은 해당하는 알고리즘을 실행하는데 걸린 시간의 평균값이다.
실험에 사용된 입력 데이터와 배정 결과는 GitHub\footnote{Dataset GitHub Link: \url{https://github.com/Seongbeom-Park/stable_marriage_one_sided_preference_data}}에서 확인할 수 있다.

\subsection{100 students \& 5 schools}
\begin{figure}[h]
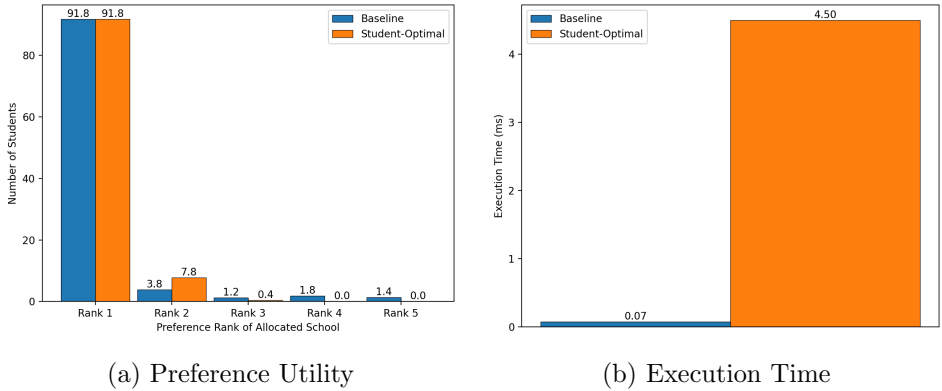

    \centering
    \begin{subfigure}{0.49\textwidth}
        \includegraphics[width=\textwidth]{./figures/student100_school5/student100_school5_utility.png}
        \caption{Preference Utility}
        \label{fig:student100_school5_utility}
    \end{subfigure}
    \hfill
    \begin{subfigure}{0.49\textwidth}
        \includegraphics[width=\textwidth]{./figures/student100_school5/student100_school5_execution_time.png}
        \caption{Execution Time}
        \label{fig:student100_school5_exec}
    \end{subfigure}
    \caption{Experiment Results with 100 Students and 5 Schools}
    \label{fig:student100_school5}
\end{figure}

Figure~\ref{fig:student100_school5_utility}는 학생 수가 100명이고 학교 수가 5개일 때, 배정 방법에 따른 선호 공리를 보여주고 있다.
모든 학교의 정원은 20명으로 설정하였다.
Student-Optimal과 Baseline 모두 가장 희망하는 학교에 배정된 인원은 평균 91.8명으로 같다.
그러나, 2지망 배정 인원은 Baseline보다 Student-Optimal이 평균 4명이 더 많다.
Student-Optimal은 3지망에 0.4명이 배정되면서 모든 학생이 3지망 이내의 학교로 배정되지만, Baseline은 3지망에 1.2명, 4지망에 1.8명, 5지망에 1.4명이 배정된다.
수행시간은 Student-Optimal이 4.50ms으로 측정되었다.

\subsection{1,000 students \& 5 schools}
\begin{figure}[h]
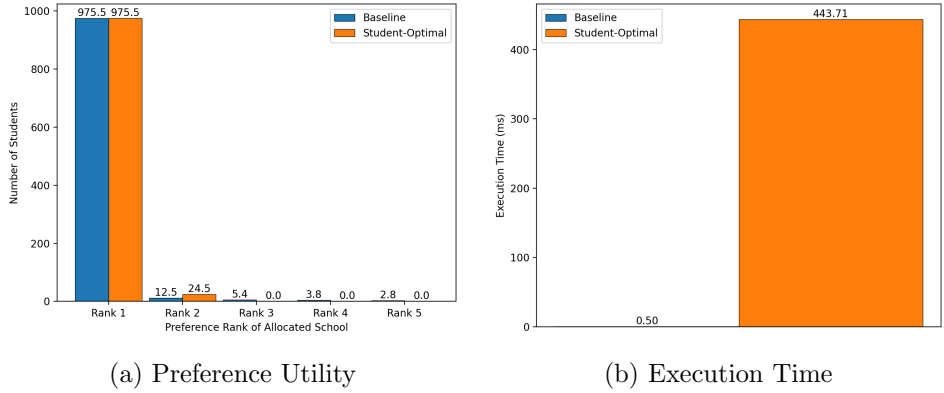

    \centering
    \begin{subfigure}{0.49\textwidth}
        \includegraphics[width=\textwidth]{./figures/student1000_school5/student1000_school5_utility.png}
        \caption{Preference Utility}
        \label{fig:student1000_school5_utility}
    \end{subfigure}
    \hfill
    \begin{subfigure}{0.49\textwidth}
        \includegraphics[width=\textwidth]{./figures/student1000_school5/student1000_school5_execution_time.png}
        \caption{Execution Time}
        \label{fig:student1000_school5_exec}
    \end{subfigure}
    \caption{Experiment Results with 1,000 Students and 5 Schools}
    \label{fig:student1000_school5}
\end{figure}

Figure~\ref{fig:student1000_school5_utility}는 학생 수가 1,000명이고 학교 수가 5개일 때, 배정 방법에 따른 선호 공리를 보여주고 있다.
학생 수가 늘어남에 따라 학교의 정원도 200명으로 증가하였다.
학교에서 받아들일 수 있는 학생 수가 늘어남에 따라, 가장 희망하는 학교에 배정되는 학생의 비율이 97.55\%로 증가하였다.
Student-Optimal은 Baseline보다 2지망에 평균 12명 더 많은 학생들을 배정하였다.
Student-Optimal은 2지망까지 모든 학생들을 배정하였지만, Baseline은 3, 4, 5지망에 각각 5.4명, 3.8명, 2.8명을 배정하였다.
수행시간은 Student-Optimal이 평균 443.71ms, Baseline은 평균 0.50ms로 측정되었다.

\subsection{10,000 students \& 50 schools}
\begin{figure}[h]
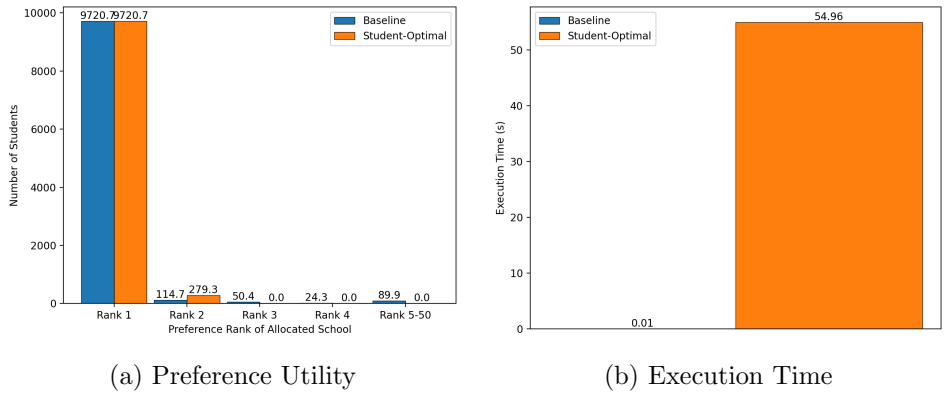

    \centering
    \begin{subfigure}{0.49\textwidth}
        \includegraphics[width=\textwidth]{./figures/student10000_school50/student10000_school50_utility.png}
        \caption{Preference Utility}
        \label{fig:student10000_school50_utility}
    \end{subfigure}
    \hfill
    \begin{subfigure}{0.49\textwidth}
        \includegraphics[width=\textwidth]{./figures/student10000_school50/student10000_school50_execution_time.png}
        \caption{Execution Time}
        \label{fig:student10000_school50_exec}
    \end{subfigure}
    \caption{Experiment Results with 10,000 Students and 50 Schools}
    \label{fig:student10000_school50}
\end{figure}

Figure~\ref{fig:student10000_school50_utility}는 학생 수가 10,000명이고 학교 수가 50개일 때, 배정 방법에 따른 선호 공리를 보여주고 있다.
모든 학교의 정원은 200명으로 동일하게 설정하였다.
가장 희망하는 학교에 배정되는 학생은 Student-Optimal과 Baseline에서 동일하게 평균 9,720.7명이 배정되었다.
Student-Optimal은 2지망에 279.3명을 배정하였고, 3지망 이하로 배정된 학생은 없다.
반면에 Baseline은 2, 3, 4지망에 각각 114.7명, 50.4명, 24.3명을 배정하였다.
Baseline은 5-50지망에 총 89.9명을 배정하였으며, 모든 지망에서 평균 0.1명 이상이 배정되었다.
수행시간은 Student-Optimal이 평균 54.96초, Baseline은 평균 9.58ms로 측정되었다.

특정 학교에 대한 선호나 기피가 없는 무작위로 생성된 데이터를 사용하였음에도 불구하고, 여러 실험 설정에서 Baseline은 지망 순위가 낮은 학교로 배정되는 학생이 발생하였다.
반면에, Student-Optimal은 학생수가 100명일 때는 3지망 이내에 모든 학생들을 배정하였고, 학생수가 1,000명, 10,000명 일때는 2지망 이내에 모든 학생들을 배정하였다.
학생 최적의 매칭은 Tie-Breaking을 통해 생성 가능한 학생 중심의 안정적인 매칭들 중 선호 공리가 최대인 매칭이기 때문이다.
비교적 높은 시간 복잡도로 인해 학생수가 증가함에 따라 수행시간이 크게 증가했지만, 멀티 스레드와 GPU 등을 활용하도록 구현하면 수행시간을 단축시킬 수 있을 것이다.

\section{Conclusion}
시험 없이 진학할 학교를 정하는 제도적 특성으로 인해 다수의 동점자가 생긴다.
동점자를 처리할 수 없는 Gale-Shapley 알고리즘\cite{gale1962college}의 특성으로 인해 추첨을 통해 Tie-Breaking\cite{irving1994stable}을 하게된다.
그러나, 무작위로 학생들의 배정 순서를 결정했을 때 많은 학생들이 선호도가 떨어지는 학교로 배정되는 문제점이 발생하고 있다\cite{abdulkadiroglu2021school, erdil2008s}.

주어진 선호도에서 생성 가능한 안정적인 매칭이 여러 개 있을 수 있기 때문에, 추첨 결과에 따라 각 지망 순위로 배정되는 학생 수가 크게 달라진다.
동점자가 있는 경우에는 최상의 품질을 갖는 매칭을 찾는 문제들은 NP의 시간 복잡도를 갖는 경우가 많다\cite{abdulkadiroglu2021school, erdil2008s, manlove2002hard, o2007algorithmic}.
얼마나 많은 학생들의 선호를 이루어 줄 수 있는지도 알기 어렵기 때문에, 생성된 매칭의 품질을 개선시키는 연구들이 진행되었다\cite{erdil2008s, irving2010finding, ravindranath2021deep}.

이 논문에서는 한 쪽은 선호도를 갖고, 다른 쪽은 정원이 제한되어 있는 상황에서, 학생 중심적이며, 안정적이고, 공리가 최대인, 학생 최적의 매칭을 찾는 방법을 소개하였다.
우선, 한 쪽만 선호도를 갖는 양면 시장에서의 학생 중심의 안정적인 매칭은 동형 선호도를 갖는 양면 시장에서 안정적인 매칭을 찾는 문제와 같음을 보였다.
또한, 선호 공리주의자(preference utilitarian)의 관점에서 공리를 정의함으로써\cite{harsanyi1977morality, singer2013companion}, 매칭의 품질을 측정할 수 있는 새로운 방법을 제시하였다.
마지막으로, 동형 선호도를 갖는 안적적인 결혼 문제(stable marriage problem with symmetric preference)를 할당 문제(assignment problem)로 환원하기 위한 가중치 변환 함수의 조건을 찾았다.

이 논문에서는 가중치의 총합이 최대인 매칭이 학생 최적의 매칭임을 증명하였다.
우선, 생성된 매칭의 안정성을 귀류법을 사용하여 증명함으로써, 학생 중심적이며 안정적인 매칭임을 증명하였다.
또한, 생성된 매칭이 선호 공리를 최대화한다는 것을 귀류법을 사용하여 증명하였다.
이를 통해 한 쪽만 선호도를 갖는 안정적인 결혼 문제에서 학생 중심적이며, 안정적이고, 공리가 최대인 학생 최적의 매칭을 다항시간 안에 찾을 수 있음을 증명하였다.

이 논문에서는 실험을 통해 배정 품질이 크게 향상됨을 정량적으로 검증하였다.
학생수를 100명에서 10,000명으로 늘려가며 무작위로 생성된 선호도 데이터로 배정 성능을 측정하였다.
Tie-Breaking으로는 선호도가 낮은 학교에도 다수의 학생들이 배정되는 데이터에서, 학생 최적의 배정은 모든 학생들을 3지망 이내로 배정하였다.
또한, 10,000명 배정에 1분 미만의 수행시간으로 충분히 실사용이 가능한 수준임을 검증하였다.

끝으로, 이 연구를 통해 더 많은 학생들에게 원하는 교육을 받을 수 있는 기회를 제공함으로써, 모두가 co-winner인 사회가 될 수 있기를 바란다.

\section*{Acknowledgement}
이 연구는 33년 동안 교육계에 헌신하신 아버지 박창호와 함께 평준화 지역의 학교 배정 제도에 대한 토론을 통해 시작하게 되었으며,
고등학교에서 학생들을 가르치고 있는 여동생 박서영에게 윤리학에 대한 도움을 받았고,
카카오에서 개발자로 활동 중인 학교 후배 김정현의 도움을 받아 증명 및 수학 기술 방법을 발전시켰다.
영문 번역은 구글 번역기의 도움을 받았다.

\medskip

\bibliographystyle{unsrt}
\bibliography{main}